 \documentclass[aps,prb,preprint,superscriptaddress]{revtex4}
\usepackage{graphicx}

\newcommand\df{\mbox{$\Delta f$}}
\newcommand\fp{\mbox{$f_{\rm pk}$}}
\def\fig#1{Fig. \ref{fig#1}}
\def\Fig#1{Figure \ref{fig#1}}

\newcommand\sxx{\mbox{ $\sigma_{xx}$}} 
\newcommand\imsxx{\mbox{Im$( \sigma_{xx} )$}} 
\newcommand\resxx{\mbox{Re$( \sigma_{xx} )$}}

\begin{document}



\title{Microwave pinning modes near Landau filling $\nu=1$ in  two-dimensional electron systems with alloy disorder }



\author{B.-H. Moon}
 \affiliation{National High Magnetic Field Laboratory, 1800 E. Paul Dirac Drive, Tallahassee, FL 32310}
 \author{L. W. Engel}
\affiliation{National High Magnetic Field Laboratory, 1800 E. Paul Dirac Drive, Tallahassee, FL 32310}
\author{D. C. Tsui}
\affiliation{Department of Electrical Engineering, Princeton University, Princeton, NJ 08544}
\author{L. N. Pfeiffer}
\affiliation{Department of Electrical Engineering, Princeton University, Princeton, NJ 08544}
\author{K. W. West}
\affiliation{Department of Electrical Engineering, Princeton University, Princeton, NJ 08544}


\date{\today}

\begin{abstract}
 
 We report measurements of   microwave spectra of two-dimensional electron systems hosted in  dilute Al alloy,  Al$_x$Ga$_{1-x}$As, for a  range of  Landau level fillings, $\nu$, around 1.   For $\nu>0.8$ or $\nu<1.2$,  the samples exhibit a microwave resonance whose frequency decreases as $\nu$ moves away from 1.  A resonance with this behavior is the signature of  solids of quasiparticles or -holes in the partially occupied Landau level, which was previously seen in ultralow disorder samples.  For $\nu<0.8$ down to as low as  $\nu=0.54$,  a resonance in the spectra is still present in the Al alloy-disordered samples, though it is  partially or  completely suppressed at $\nu=3/5$ and $1/2$, and is strongly damped over much of this $\nu$ range.     The resonance also shows a striking enhancement in  peak frequency for $\nu$ just below 3/4.      We discuss possible explanations of the   resonance behavior for $\nu<0.8$  in terms of  the  composite fermion picture.

\end{abstract}


\maketitle



In two-dimensional electron systems (2DES)  in high magnetic field,  $B$, 
Wigner solids\cite{lozo,lamgirvin,kunwc,narevichfertig,jaincfwc,archer1,archer2,reentrant,vgwc, buhmann,kukushkinwctri,ee,williams91,willett1,willett2,lessc,clibdens,yewc,murthyrvw,msreview}   pinned by residual disorder occur  for sufficiently small Landau level filling factor, $\nu$.  At the low $\nu$ termination of the fractional quantum Hall effect (FQHE) series, such a solid can  appear  in dc transport as an insulating phase.   
 By means of microwave spectroscopy like that employed to obtain the data in this paper,   Wigner  solids of  dilute quasiparticles or quasiholes in the presence of one or more filled Landau levels have also been found\cite{yongiqhwc,hanskyrme,ruperti3}, in  extremely high mobility,  low disorder 2DES.  
 This type of solid has been named an integer quantum Hall effect   Wigner solid (IQHEWS) since the vanishing diagonal conductivity in the wings of the integer quantum Hall effect (IQHE) diagonal conductivity minimum is  due to the pinning of the  solid.  An analogous solid composed of quasiparticles and -holes of the 1/3 fractional quantum Hall liquid has also been reported\cite{hanfqhe}. 

Microwave spectroscopy is known to directly measure pinned solids in high magnetic field, since these  solids 
  exhibit    resonances in their microwave spectra \cite{ ee,williams91,yewc,yongiqhwc,bubres,ruperti3,hanfqhe,hanskyrme}.    The resonances are understood as pinning modes\cite{williams91,willett1,willett2,lessc,clibdens,yewc,flr}, in which pieces of the  solid oscillate within the disorder potential.   The frequency, \fp, of the pinning mode is determined by the disorder in the sample and by the stiffness of the solid.  \fp\ always increases when the disorder strength is increased, but in weak pinning theory\cite{flr,chitra,fertig,foglerhuse}  
when the solid is made stiffer, \fp\ decreases.  This behavior, verified experimentally\cite{clibdens,yewc,yongiqhwc},  occurs because for a stiffer crystal the carrier is less closely associated with 
   disorder potential features.  In the case  of the 
IQHEWS\cite{yongiqhwc,hanskyrme}, near $\nu=1$  the quasiparticle or -hole density is $n^*=|\nu^*|n/\nu$ where $\nu^*=\nu-1$,  and $n$ is the carrier density of the 2DES. Hence the generic behavior of the IQHEWS is that \fp\ is largest near the integer $\nu$, and the resonance initially increases in strength as  one moves away from integer $\nu$ and   $|\nu^*|$ increases. 
  At still larger $|\nu^*|$, above 0.15  in low disorder samples\cite{yongiqhwc} near $\nu=1$, the resonance fades away as the carriers become  dense enough for the solid to melt. 

 Earlier reports of IQHEWS have been on quantum well samples of the highest quality.    These include\cite{hanskyrme} a 50 nm well, with mobility 
$\mu=\approx15  \times 10^6$ cm$^2$/V-s, and a 30 nm well with $\mu \approx27  \times 10^6$ cm$^2$/V-s.  Here we report   IQHEWS, identified from  microwave spectroscopy, in samples with  far lower mobility,  $1.4$ and $2.3\times 10^6$ cm$^2$/V-s.  The   samples used in this work contain the 2DES in dilute Al alloy,  Al$_x$Ga$_{1-x}$As, so  the isoelectronic Al impurities provide an additional  component of the overall disorder.  The length scales  of this scattering potential are much less than the magnetic length, $l_B$,  under all conditions in this paper.

This paper shows  that even with the Al alloy disorder, for    $|\nu^*| < 0.2$  the resonances mainly follow the behavior expected of an IQHEWS resonance, with \fp\ increasing as $\nu=1$ is approached.      \fp\ vs $\nu$, also shows inflections likely due to skyrmion crystallization\cite{hanskyrme}.   Furthermore, more detailed study of the pinning resonance  also shows that
relative to samples without Al, the   resonance range is greatly extended  on the hole side ($\nu < 1$)  down to $\nu\sim0.6$. In this extended range,    the behavior of the resonances is more complex than was reported\cite{yongiqhwc} in Al-free samples, and the interpretation 
may  require phase transitions within the solid.   As $\nu$ decreases below 0.8,   
the trend of decreasing \fp\ reverses, and \fp\  vs $\nu$     develops   a  strongly disorder-dependent maximum for  $\nu$ just below $3/4$.   The resonance  also is suppressed or partly suppressed at    $\nu=2/3$ and $3/5$.    Particularly away from $\nu=1$ and from the maximum in \fp\ vs $\nu$, the resonance is highly damped with a rapid rise at low frequency, $f$, in  real diagonal conductivity, $\resxx(f)$.   We observed such a highly damped resonance  in the ranges of $\nu$  between $2/3$ and $3/5$ reported\cite{wanlireentrant}   to exhibit insulating behavior in the partially filled Landau level  in dc transport

We  present data from two samples with Al$_x$Ga$_{1-x}$As  channels.  Samples from the same wafers were used in earlier work in dc transport 
\cite{wanliscatt,wanlisclg,wanlisclganderson,wanlisclgcross,wanlireentrant}.  Characterization\cite{wanliscatt} 
of the samples at $B=0$ showed that the spatial distribution of the Al is random.        Both samples are single heterojunctions  with the 2DES approximately 180 nm below the top surface of the sample.    One sample has $x=0.33$\%, density $n\approx 2.1\times 10^{11}$ cm$^{-2}$ and as cooled mobility $\mu$ about $2.3 \times 10^6$ cm$^2$/V-s; the other sample has $x=0.85$\%, $n\approx 2.5\times 10^{11}$ cm$^{-2}$ and   $\mu\approx1.4  \times 10^6$ cm$^2$/V-s.  A previous paper\cite{byounglonu} showed that samples from this series exhibit pinning mode resonances at the low $\nu$ termination of the FQHE series, which for these disordered samples occurred for $\nu$ below that of the 1/3 FQHE.  The presence of disorder is known to raise the $\nu$ at which the FQHE series is terminated in an insulator.    At least for samples grown in the same series, for which the component of    disorder not due to the Al was the same,    additional Al  raised  \fp, as one would expect for a pinning mode. 

 Microwave spectra of the 
complex diagonal conductivity \sxx, were obtained from  the propagation characteristics  of a transmission line\cite{ee} that is capacitively coupled to the 2DES.  The transmission line  we used is shown schematically in \fig{bsc}a.    The lines  are of the coplanar waveguide (CPW)  type \cite{clibdens,yewc,yongiqhwc,bubres,ruperti3,hanfqhe,hanskyrme,zhwimbal}, with a center conductor  separated from two broad ground planes by slots of width $W=30\ \mu$m, and are patterned directly onto the top of the sample.    We present complex diagonal conductivity
calculated from   the    model  of  ref. \onlinecite{zhwimbal} for  the CPW coupled to the 2DES.     The results are within about  25\% of those from  the high frequency low loss approximation,
  $  \sigma _{xx} \approx -W|\ln(s  )|/ Z_0d$,    where $s$ is  a complex transmission coefficient,     $Z_0=50\ \Omega$ is the characteristic impedance calculated from the transmission line geometry for $\sigma_{xx}=0$, and $d$ is the length of the coplanar waveguide  ($d=28$ mm).     $s$ is normalized to the transmission at $\nu=1$, at which \sxx\ vanishes due to the IQHE.    For all measurements, the   2DES temperature was approximately 60 mK, as read by nearby resistance thermometers.   Slight broadening of the resonance could be discerned on increasing the  temperature above that value.  We verify that the  data were taken in the low microwave power limit, in which further decrease of power   does not affect the measured spectrum.
  
\Fig{bsc}b shows \resxx\ vs $B$ for the two samples, both taken at 100 MHz.    \resxx\ for the  $x=0.33\%$ sample shows a
 broad minimum  at the $\nu=1 $ IQHE, and FQHE minima at $3/5, 2/3, 4/3, 7/5,  8/5$ and $ 5/3$.   The curve is composed of points read off from spectra taken at discrete $B$, and the density of points gives the traces an artifactual jagged appearance  near $\nu=3/5$ and $2/3$. The width of the IQHE minimum is about 0.3 in $\nu$, measured from halfway up the rises on either side.   The trace from the $x=0.85\% $ sample shows a much wider IQHE around $\nu=1$, with $\nu$-width measured the same way of 0.56.  The FQHE at 2/3 or 3/5 is not  apparent,  though there are minima near 4/3 and 5/3.    For both traces, the $\nu=2$ IQHE minimum  is well-developed but appears at \resxx\ slightly elevated from that  at  $\nu=1$; we ascribe this to  weak, $B$-dependent parallel conduction in the wafers.     

 \Fig{colspx}a  is an image plot of   spectra,  \resxx\ vs $f$,  for $\nu$ between 0.6  and 1.4 for the $x=0.33\%$ sample. The $x=0.33\%$ trace in    \fig{bsc}b is a 0.1 GHz cut of this image.  There are clear resonances on each side on $\nu=1$, which decrease in \fp\ as $\nu$ moves farther from 1.     
 As $\nu$ goes above 1.2, the resonance fades into a flat spectrum. 
In contrast, as $\nu$ goes below $0.8$,   \resxx\ vs $f$ continue  to exhibit maxima.  

  \resxx\ spectra for all $\nu\le1$ and $x=0.33\%$ are shown  in \fig{x33spx}a.  For $\nu$ between 0.8 and 2/3, \resxx\ shows a strongly asymmetric peak, with rapid rise at low $f$ and a high $f$ tail, whose extent is $\nu$ dependent.   At 2/3 the  spectrum becomes flat. The asymmetric peak  reappears for $\nu<2/3$, but weakens for the lowest $\nu$ trace at 0.61. 

The color image in \fig{colspx}b gives an overview of the  \resxx\ spectra for the $x=0.85\%$ sample, and the spectra for $\nu\le1 $  are plotted as curves in 
\fig{x85spx}a.
Between $\nu= 1.05$ and $1.2$, and also between between $\nu= 0.94$ and $0.8$  there is a resonance whose \fp\ increases  as $\nu=1$ is approached.   
As $\nu$ goes below 0.8, the  resonance moves to higher \fp, 
with a maximum \fp\ around $\nu=0.74$.  As 2/3 is approached from higher $\nu$ the resonance decreases in frequency and gradually becomes asymmetric.  A resonance is present at  2/3 and 3/5, which lie within the low-$f$ IQHE plateau,  but these rational fractional $\nu$ are easily distinguished in \fig{colspx}b by their reduced overall \resxx.  
 
\begin{figure}
 \includegraphics[width=3in]{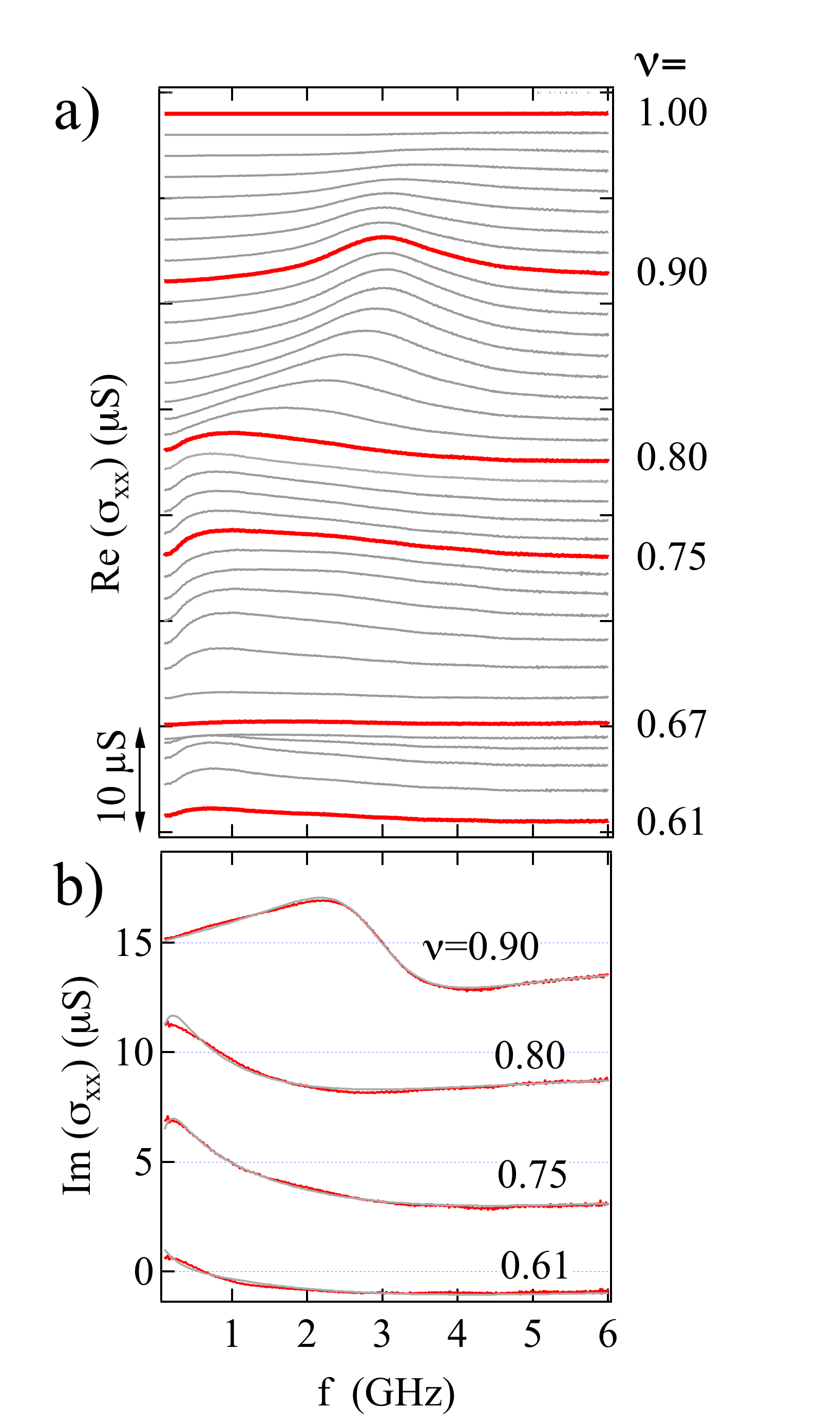}
 \caption{Spectra for the sample with $x=0.33\%$ for different Landau level fillings $\nu$.  a) \resxx\ vs frequency, $f$.  Successive spectra were taken  at $\nu$-intervals of   $0.0107$, and  are offset vertically proportionally to $\nu$. 
 Landau fillings of  certain  spectra (red)  are marked at right.  b) \imsxx\ vs $f$ for  several $\nu$. Successive spectra are offset by 5 $\mu$S.    Fits are light lines, data are darker (red) lines.    }\label{figx33spx}
 \end{figure}

\begin{figure}
 \includegraphics[width=3in]{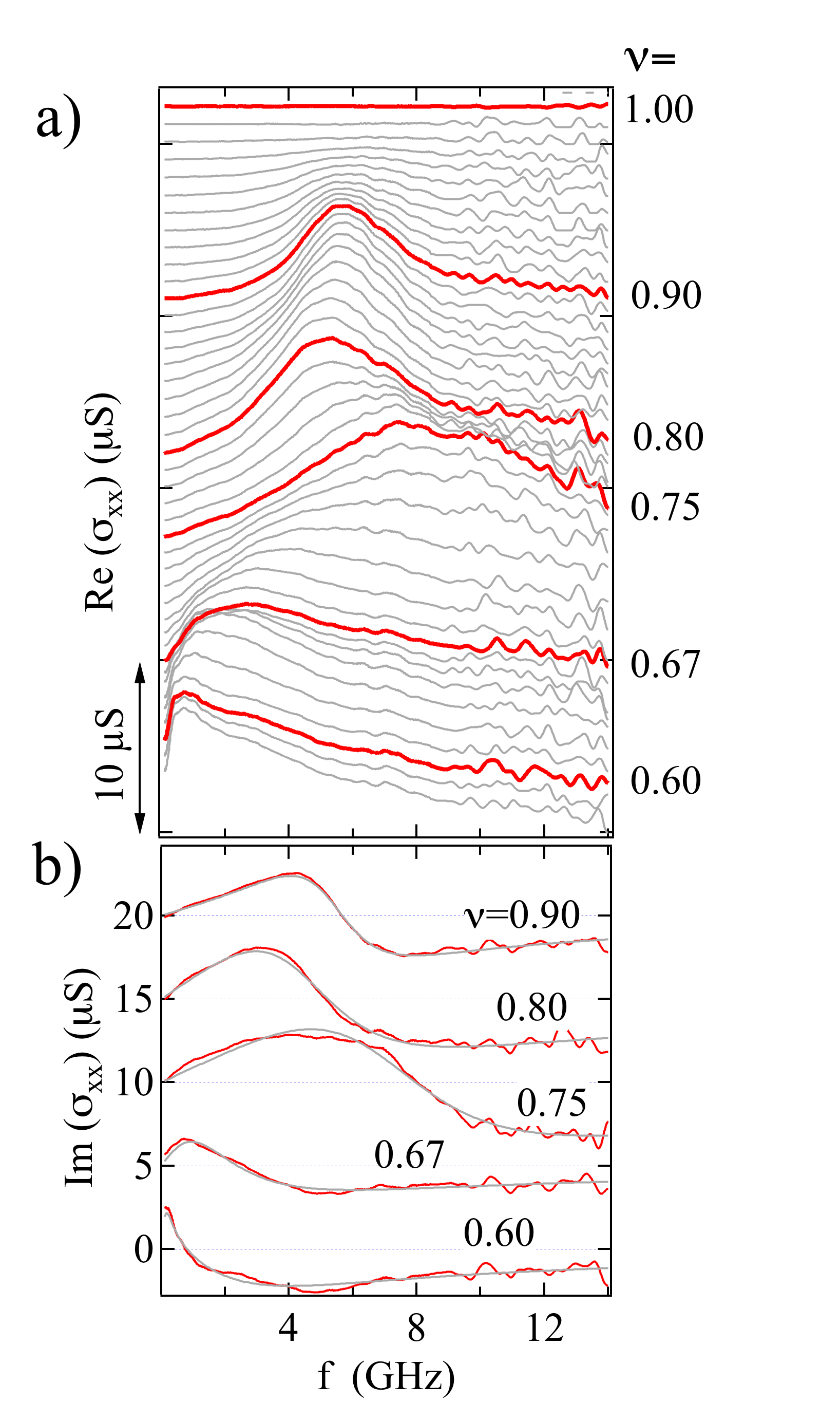}
 \caption{a)    Spectra for the sample with $x=0.85\%$ for different Landau level fillings $\nu$.  a) \resxx\ vs frequency, $f$.  Successive spectra were taken  at $\nu$-intervals of   $0.0111$, and  are offset vertically proportionally to $\nu$. 
 Landau fillings of  certain  spectra (red on line)  are marked at right.  b) \imsxx\ vs $f$ for  several $\nu$. Successive spectra are offset by 5 $\mu$S.     Fits are light lines, data are darker (red on line) lines.     }\label{figx85spx}
 \end{figure}

 To characterize the spectra  it is useful to perform fits to extract resonance parameters, particularly when the resonances are asymmetric and broad.  
 Figures \ref{figx33spx}b and \ref{figx85spx}b show \imsxx\ vs $f$ at several fillings as marked.   The \imsxx\ spectra can be fit   to the imaginary part of the response of a  Drude-Lorentz harmonic oscillator \cite{aandm}, 
 $\sxx(f) =  \{ \sigma_0  f \df /[   f \df +i  ( f^2-\fp^2)      ]\}$, a form commonly used \cite{aandm} for dielectrics,     where the  parameters  $\sigma_0$ and \df\ are  the amplitude  and  line width.   
  The  corresponding real part, with the same fit parameters,  fit the \resxx\ spectra to within   frequency-independent additive constants. 
The fit, common for lossy dielectrics \cite{aandm}  was chosen empirically, and is not readily connected to  theoretically derived forms\cite{chitra}  of pinning modes for weak disorder in high $B$, possibly since these existing theories are specific to the weak disorder limit.

\begin{figure}
 \includegraphics[width=3in]{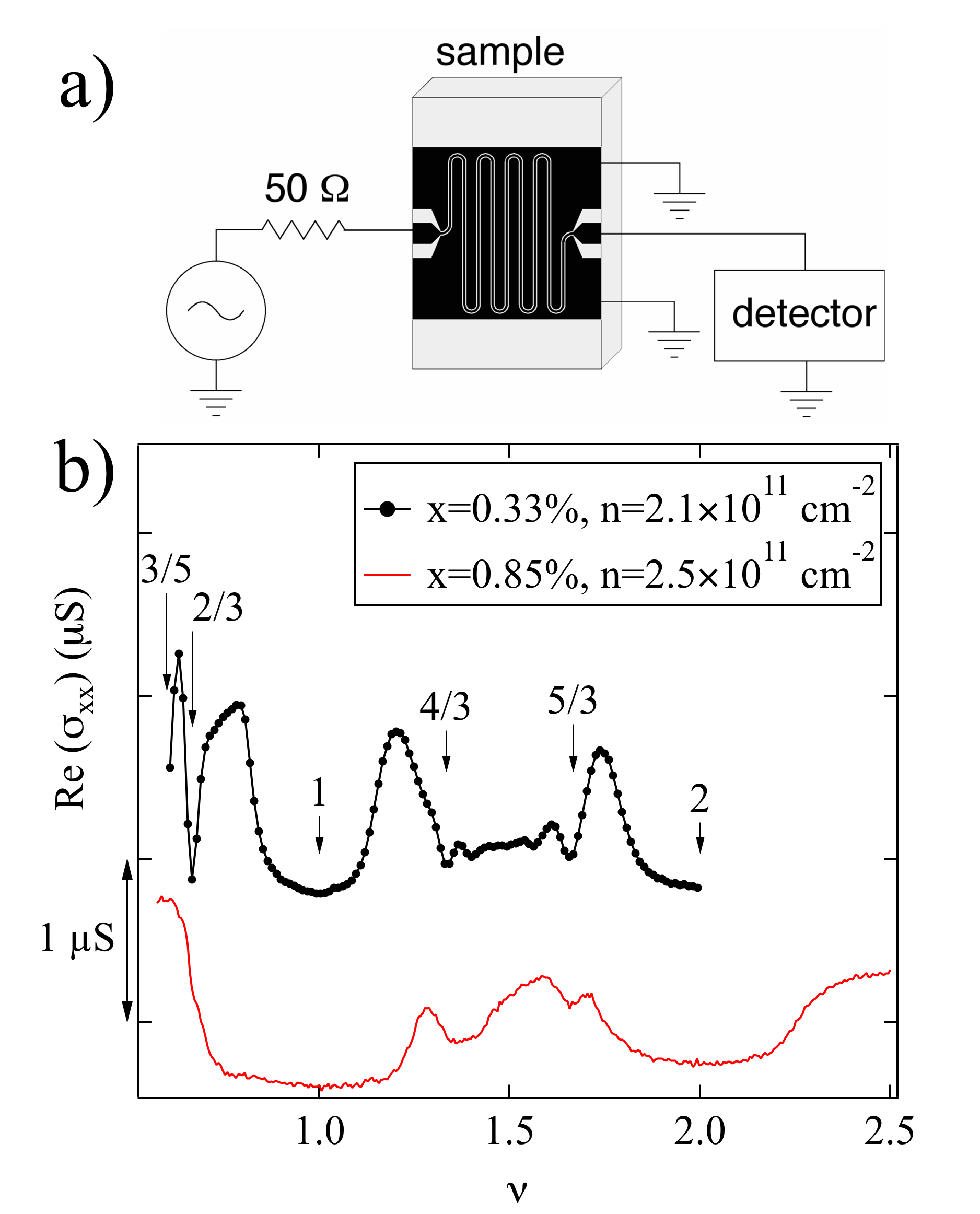}
 \caption{a)  Schematic representation of the microwave circuit used in our measurement, not to scale.    Black areas represent metal films on the sample surface. \ \ b)   Real part of diagonal conductivity \resxx, vs Landau filling factor $\nu$ for the two samples at 0.1 GHz.  The upper, 
 ($x=0.33\%$) curve was  interpolated from spectra, and has a lower density of data points, which are shown as dots.     
   }\label{figbsc}
 \end{figure}
\begin{figure}
 \includegraphics[width=3in]{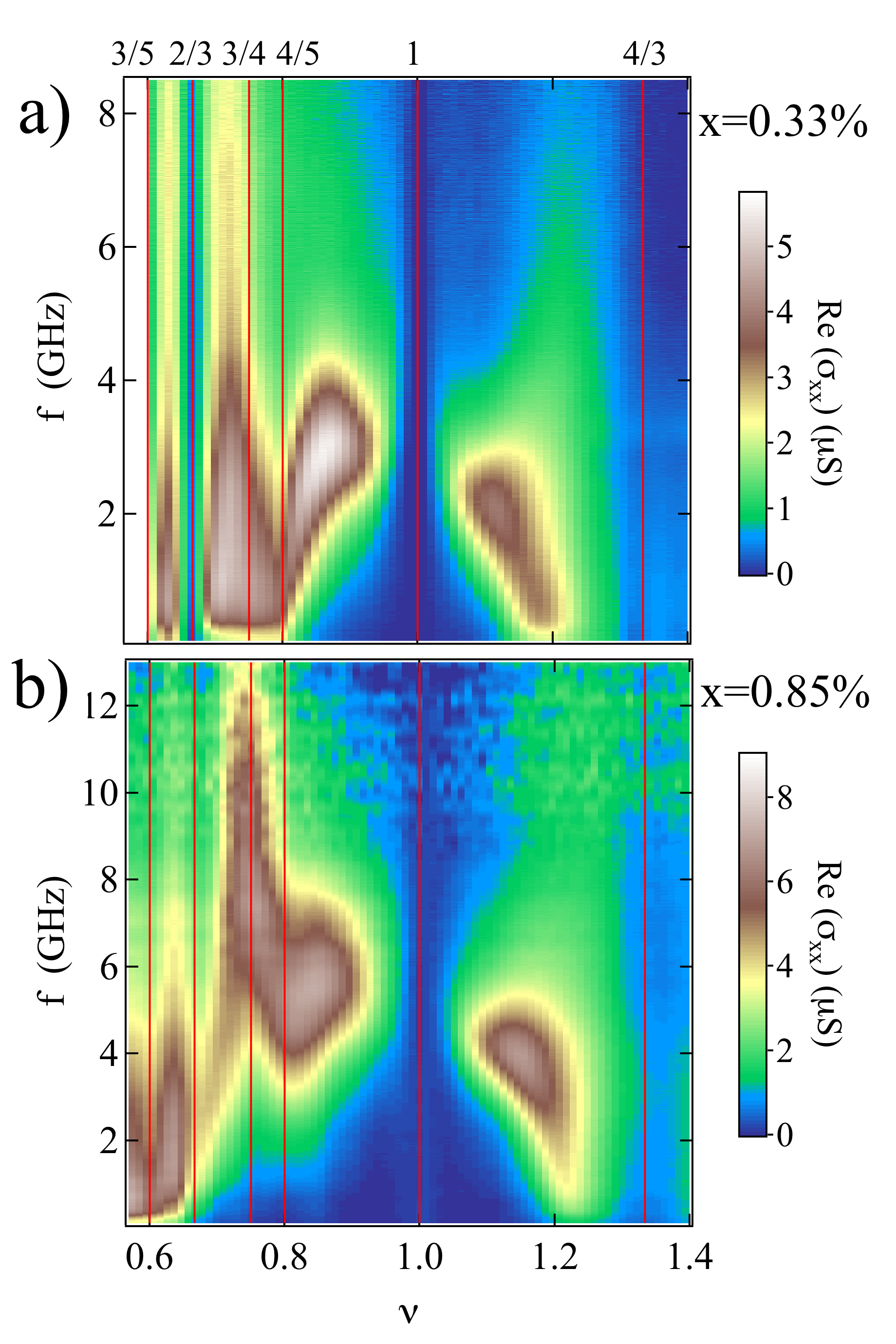}
 \caption{Color-scale plots showing \resxx\ in $f$-$\nu$ plane a)  $x=0.33\%$   \ \ b) $x=0.85\%$   Red vertical lines
 mark   Landau filling factors  $3/5,2/3,3/4,1$ and $4/3$. }\label{figcolspx}
 \end{figure}
\begin{figure}
 \includegraphics[width=3in]{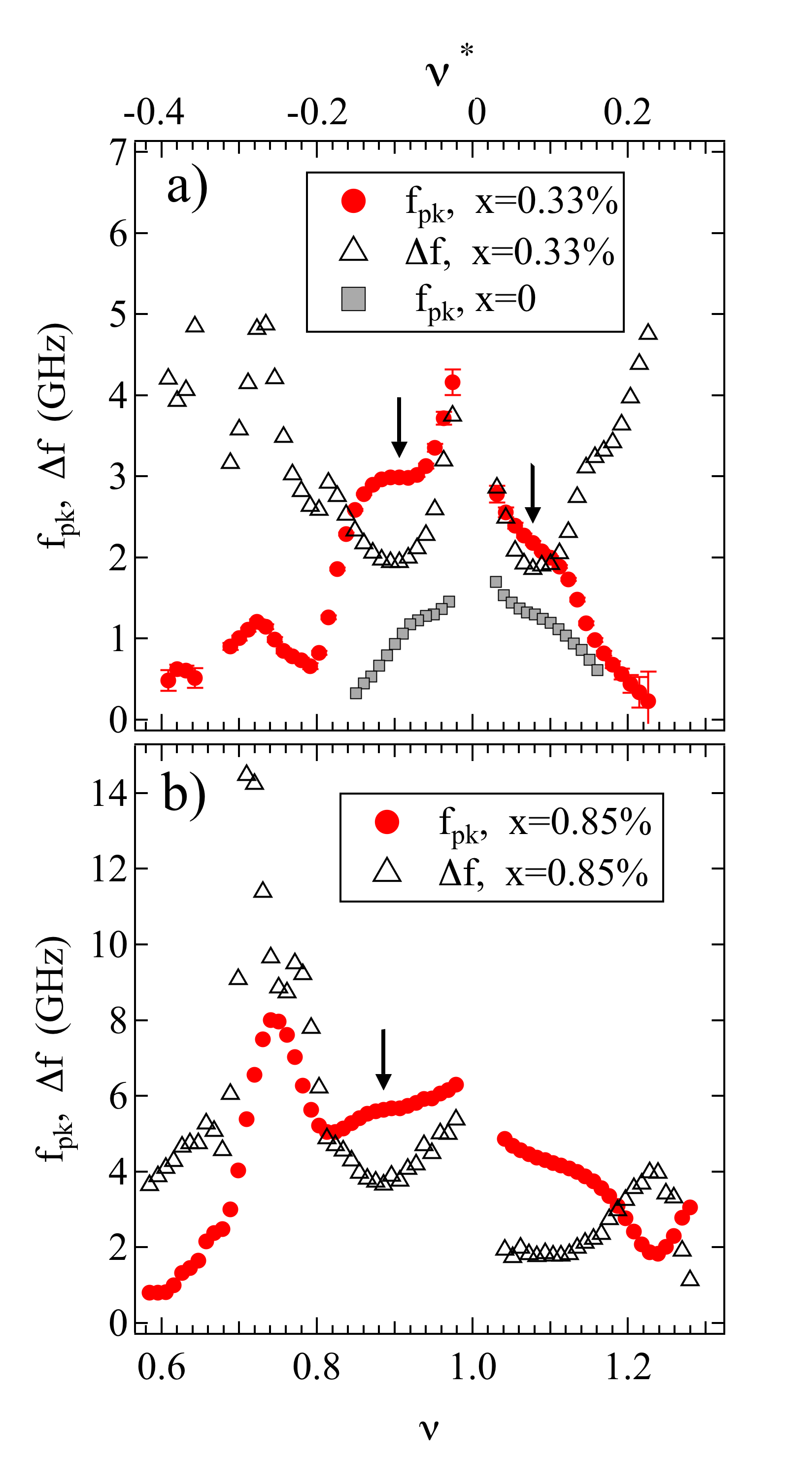}
 \caption{a)   Resonance parameters peak frequency, (\fp) and linewidth, (\df) vs Landau level filling $\nu$.  $\nu^*=\nu-1$ is shown on  the top axis.  for a) $x=0.33\% $,  $x=0$ data are  \fp\ from ref. onlinecite{hanskyrme}, for a 30 nm quantum well  with   mobility $27\times 10^6$ m$^2$/V-s.  b)  $x=0.85\%$. Arrows mark  the inflections discussed in the text.  }\label{figsum}
 \end{figure}

\Fig{sum} shows \fp\ and the linewidth, \df, vs $\nu$ and $\nu^*$ for both the samples.   Both \fp\ and \df\ come from fits of \imsxx\ vs $f$, like those shown in \fig{x33spx} and \fig{x85spx}.  \Fig{sum}a also shows \fp\ vs $\nu$  from ref. \onlinecite{hanskyrme}, on an ultralow disorder quantum well sample, with $x=0$, $n=2.7 \times 10^{11}$ cm$^{-2}$ and  $\mu\approx27\times 10^6$ m$^2$/V-s.    The $x=0.33 $ and $0.85\%$ \fp\  traces are decreasing as $\nu$ moves away from 1, out to $|\nu^*|\approx 0.2$.   \fp\ vs $\nu$  for 
the $x=0$ ultralow disorder sample  likewise decreases on moving away from 1, though the range of $\nu$ for  that data is less. 
Inflections    are marked with arrows for  the  \fp\ vs $\nu$  traces for $x=0.33\%$ and for $x=0.85\%$, but are also visible in 
  the $x=0$ data, for slightly smaller $|\nu^*|$.
Inflections are marked by down arrows for the $x=0.33\%$ \fp\ vs $\nu$ data in  \fig{sum}a, at $\nu=0.91$ and $1.08$.  Inflections are also visible in   the $x=0$  \fp\ vs $\nu$, for slightly smaller $|\nu^*|$.  A similar but much weaker feature in    $\nu\approx 0.88$ is marked in \fig{sum}b as well.

The main point of this paper is that the behavior of the resonance for $|\nu^*|<0.2$, with monotonic decrease of \fp\  as $\nu^*$ is evidence for an IQHEWS, much like that found  in 
ultralow disorder samples. The decrease in \fp\ is due to the increase in $n^*$, which results in a stiffer solid that  avoids features of the disorder which increase pinning.     \fp\ is larger than the ultralow disorder sample, as expected for larger disorder, but the disorder, possibly because of its homogenous distribution and short range, still allows IQHEWS formation.  

The inflections in the \fp\ vs $\nu$ traces of \fig{sum} are also similar to effects in the $x=0$ ultralow disorder sample. 
   The inflections can be  regarded as due to a sudden decrease in the pinning as $\nu$ approaches 1, superposed on  the pinning increase  engendered  by  the reduction in quasiparticle density.  
Based on ref.   \onlinecite{hanskyrme}  we interpret this as due to formation of skyrmions\cite{barret,sondhi,schmeller}  containing multiple   spins.     Crystals\cite{bayotmelinte,fertigsk} of such skyrmions  can form near  $\nu=1$ state when 1) $\nu$ sufficiently close to 1 and 2) when Zeeman energy is sufficiently small relative to Coulomb energy, that is, for  sufficiently small $\tilde{g}=g\mu_BB_T/(e^2/4\pi\epsilon\epsilon_0l_B)$,where $B$ is total magnetic field.
 For the present samples $\tilde{g}=  0.017, 0.019$ respectively for  the $x=0.33,0.85\% $, using $|g|=0.44$.  
The sample that produced the $x=0$ trace of \fig{sum}a   had $\tilde{g}=0.019$ in perpendicular magnetic field;   the inflections  in \fp\ vs $\nu$ of this sample were shown\cite{hanskyrme} to disappear on applying in-plane magnetic field to increase $\tilde{g}$.  In-plane field experiments on the alloy-disordered samples    would verify the skyrmion based interpretation.  
  The disorder at the $ x=0.33\%$ level   actually appears to  make the skyrmion effect  {\em more prominent}   than in the $x=0$ ultralow disorder sample.   A skyrmion effect 
 in the Al disordered sample also indicates a remarkable robustness of this   interaction-driven phenomenon in the presence of 
 homogenous short-range disorder.

As $\nu$ goes below $0.8$  (and also above 1.2 in a small range for $x=0.85\%$),
   the decrease of \fp\ with $|\nu^*|$ reverses, and the interpretation of the resonance is less clear. 
For $\nu<0.8$,    \fp\ increases with decreasing $\nu$ and develops a peak around $0.72$ for $x=0.33\%$, and a much more pronounced peak around $\nu=0.74$ for $x=0.85\%$.  
The behavior seen for $\nu<0.8$ is clearly enabled by the disorder, since the disorder extends the resonant range into that region. 
Disorder stabilizes a solid  vs an FQHE liquid since the solid  optimizes its energy in the disorder potential, gaining its 
  pinning energy\cite{pricezhulouie}, and the transition to insulator at the low $\nu$ termination of the FQHE series occurs around $\nu=1/3$ in Al alloy-disordered samples\cite{wanlireentrant,byounglonu}.

Except for a small range near $\nu=1.2$ for $x=0.85\%$, the  overdamped ($\df > \fp$ on \fig{sum}) regions  have $|\nu^*|>0.2$. 
The  most strongly  overdamped resonance spectra are for $\nu$ well below $1$, for  $\nu<0.8$ for $x=0.33\%$ and   $\nu<2/3$ for $x=0.85\%$.     Their spectra are characterized by a sharp rise in \resxx\ at low $f$.  In the $\nu$ regions where these spectra occur, the  brown color in \fig{colspx} comes to within  0.1 GHz of the lower axis.   The spectra are qualitatively different from  the better-developed resonances seen   closer to integer $\nu$, or in the low $\nu$ insulator,   both in the present  samples \cite{byounglonu} and in low-disorder Al-free quantum wells \cite{yongiqhwc,hanskyrme}.    

Earlier dc transport measurements \cite{wanlireentrant}  particularly on a sample from the same $x=0.85\%$ wafer   showed a reentrant integer quantum Hall effect (RIQHE), between the $ 2/3$ and $3/5$ FQHEs.  The  RIQHE  has vanishing diagonal resistance and Hall resistance quantized to $h/e^2$, and indicates that quasiholes of the partially occupied Landau level are insulating.  In the RIQHE $\nu$ range of ref. \onlinecite{wanlireentrant}  we find resonances of the strongly overdamped type in both samples.  Possibly because of sample 
processing, variation within the wafer, or cooldown procedure, we do not find a 2/3 FQHE in the present  $x=0.85\%$ sample.   There is a 2/3 FQHE in  the present $x=0.33\%$ sample,    so in that sample the  insulator between 2/3 and 3/5 is reentrant.   
 
 While the  interpretation of the resonance  for $|\nu^*|>0.2$ is unclear, we present different possibilities, involving the composite fermion  (CF) picture \cite{jainbook}.  At  rational fractional $\nu$ of FQHE states  the resonance can be suppressed, hence it is reasonable to assume that the CF picture,  which has explained many aspects of the FQHE, will be of value even in the presence of the disorder potential. 
    There has been recent detailed theoretical treatment 
  \cite{archer1,archer2} of composite fermion (CF) Wigner crystals  in disorder-free systems.  A CF Wigner crystal,  denoted $^{2p}$CFWC, is composed of CFs with   $2p$     flux bound to an electron.    The theory, presented in ref. \onlinecite{archer2} for $\nu<4/17$, predicts a series of   transitions between different $^{2p}$CFWC with $2p$ increasing as $\nu$ decreases, in particular a change from a     $^{2}$CFWC to  $^{4}$CFWC  as $\nu$ decreases from 1/5 to 1/6.    Since the disorder extends $\nu^*$ for the resonance well beyond where solid in a low-disorder system is expected, the theory does not cover the presently observed resonance range.   It may be that   CF vortex number transitions, if present in these samples, are strongly affected by the disorder, so that the $\nu^*$ where it occurs is different than predicted.       
  
One of the most striking features of the data are the peaks around $\nu=0.72$ to $0.74$, not far 
from $\nu=3/4$ where $2p=4$ CFs form a Fermi surface, and move in a zero effective magnetic field.  It may be that this region 
is characterized  by a CF solid more like that considered for $B=0$ \cite{chitragiamarchiB0},  in which the particle size  is determined by
competition between kinetic energy and interaction; such a solid would need to be  stabilized by the dense homogenous disorder.

 One possible cause of the large damping would be pockets of FQH liquid within the insulating phase.  It is unlikely the  pockets are localized in the Al disorder, though it may be that  the much weaker,  large length scale disorder associated with the remote ionized donors is containing liquid pockets, as described in ref. \onlinecite{yac}.  An inhomogeneous  microemulsion phase of the sort proposed in 
refs. \onlinecite{jameikivspiv,kivspiv} is also conceivable.   If there are pockets of CF liquid, the additional damping for $\nu\sim 3/4 $  
is explainable as due to dissipation in the liquid pockets. The increase of \fp\ near that frequency could involve  the pockets to be  screening the Coulomb interaction  between carriers of the solid, reducing its effective stiffness at long range.

In summary we have found that Al-alloy disordered samples exhibit  the signature of an IQHEWS for $|\nu^*|$ below about 0.2. The IQHEWS is remarkably similar to that in ultralow disorder samples, and shows some likely effects of skyrmions. 
For $\nu<0.8$, where resonances are not seen for ultralow disorder samples, 
 the resonance has  an  evolution  that is more complex than that seen in low disorder samples.  \fp\ increases with decreasing $\nu$  as $\nu$ goes below $0.8$, and  a maximum in \fp\ vs $\nu$  occurs for $\nu$  just below $3/4$.  The CF picture provides possible explanations of the features.   The disorder extends the $\nu$ range of the existence of the resonance, so that there is reentrance at least of a highly damped resonance, around FQHE states.

We thank Kun Yang and J. K. Jain for helpful discussions and A. T. Hatke for his comments on the manuscript.   The work
at Princeton was funded through the NSF through MRSEC DMR-0819860 and the Keck
Foundation and the Gordon and Betty Moore Foundation (grant GBMF4420).
The microwave spectroscopy work at NHMFL was supported  through DOE grant DE-FG02-05-ER46212 at NHMFL/FSU. NHMFL is supported by NSF Cooperative Agreement No. DMR-0084173, the State of Florida and the DOE.



\begin{thebibliography}{}
\bibitem{lozo}Y. E. Lozovik and V. I. Yudson, 
  JETP Lett.,  {\bf 22}, 11 (1975).
  \bibitem{lamgirvin}
 P. K. Lam and S. M. Girvin,    
  Phys. Rev. B {\bf 30}, 473 (1984).
\bibitem{kunwc}
   Kun Yang, F. D. M. Haldane, and E. H. Rezayi
Phys. Rev. B {\bf 64}, 081301 (2001).
\bibitem{narevichfertig}R. Narevich, Ganpathy Murthy, and H. A. Fertig,  
Phys. Rev. B {\bf 64}, 245326 (2001).

\bibitem{jaincfwc}Chia-Chen Chang, Csaba T\"{o}ke, Gun Sang Jeon, and Jainendra K. Jain 
Phys. Rev. B {\bf 73}, 155323 (2006);  Chia-Chen Chang, Gun Sang Jeon, and Jainendra K. Jain 
Phys. Rev. Lett.{\bf 94}, 016809 (2005). 
\bibitem{archer1}	  Alexander C. Archer and Jainendra K. Jain
Phys. Rev. B 84, 115139 (2011).
 \bibitem{archer2}A. C. Archer, Kwon Park, Jainendra K. Jain, Phys. Rev. Lett., {\bf 111}, 146804 (2013).
 
 
 
 
 \bibitem {reentrant}
H.~W. Jiang, R.~L. Willett, H.~L. Stormer, D.~C. Tsui, L.~N. Pfeiffer, and
  K.~W. West,  {Phys. Rev. Lett.}, {\bf 65}, 633  (1990).
 \bibitem{vgwc}
V.~J. Goldman, M~Santos, M~Shayegan, and J.~E. Cunningham,
 Phys. Rev. Lett.  {\bf 65}, 2189 (1990).


   \bibitem{ee}E. Y. Andrei, G. Deville, D. C. Glattli, F. I. B.
Williams. E. Paris, and B. Etienne, 
{ Phys. Rev. Lett.} {\bf 60}, 2765 (1988).

 



\bibitem{buhmann}H. Buhmann, W. Joss, K. v. Klitzing, I. V.
Kukushkin, A. S. Plaut, G. Martinez, K. Ploog, and V. B. Timofeev, 
 Phys. Rev. Lett. {\bf 66},  926 (1991).
\bibitem{kukushkinwctri}I. V. Kukushkin, Vladimir I. FalÕko, R. J. Haug, K. von Klitzing, K. Eberl, and K. Tštemayer
Phys. Rev. Lett. {\bf 72}, 3594(1994). 
  \bibitem{williams91} F. I. B. Williams, P. A. Wright, R. G. Clark, E. Y. Andrei, G. Deville, D. C. Glattli, O. Probst, B. Etienne, C. Dorin, C. T. Foxon, and J. J. Harris, 
Phys. Rev. Lett. {\bf 66}, 3285 (1991).
 \bibitem{willett1} M. A. Paalanen, R. L. Willett, R. R. Ruel, P. B.
Littlewood, K. W. West, L. N. Pfeiffer and D. J. Bishop, 
Phys. Rev. B {\bf 45}, 11342 (1992).
  \bibitem{willett2} M. A. Paalanen, R. L. Willett, R. R. Ruel, P. B.
Littlewood, K. W. West, L. N. Pfeiffer and D. J. Bishop, 
\prb{\bf 45}, 13784 (1992).

 \bibitem{lessc}  L. W. Engel, C.-C. Li, D. Shahar, D. C. Tsui and 
 M. Shayegan,   
 Solid State Commun., {\bf 104} 167-171  (1997).
 \bibitem{clibdens} C.-C. Li, J. Yoon,L. W. Engel  D. Shahar, D. C. Tsui  and M.
Shayegan,  
Phys. Rev.   B  {\bf 61},  10905 (2000).
\bibitem{yewc} P. D. Ye, L. W. Engel, D. C. Tsui, R. M. Lewis, L. N. Pfeiffer, and K. West, Phys. Rev. Lett. {\bf 89}, 176802 (2002).
  \bibitem{murthyrvw}  G. Sambandamurthy,  Zhihai Wang,  R. M. Lewis, Yong P. Chen, L. W. Engel,
D. C. Tsui, L. N. Pfeiffer  and  K. W. West,
 Solid State Commun. {\bf 140}, 100  (2006)  contains a brief review. 
\bibitem{msreview} M. Shayegan, in Perspectives in Quantum Hall 
Effects, edited by S. Das Sarma and A. Pinczuk (Wiley-Interscience, New York, 1997), p. 343.

\bibitem{yongiqhwc} Yong Chen, R. M. Lewis ,  L. W. Engel , D. C. Tsui , P. D. Ye , L. N. Pfeiffer and K. W. West,  Phys. Rev. Lett. \textbf{91}, 016801 (2003). 

    \bibitem{hanskyrme} Han Zhu, G. Sambandamurthy, Yong P. Chen, P. Jiang, L. W. Engel, D. C. Tsui, L. N. Pfeiffer, and K. W. West, 
  Phys. Rev. Lett. {\bf 104}, 226801 (2010).
     \bibitem{ruperti3} R. M. Lewis, Yong Chen, L. W. Engel, D. C. Tsui, P. D. Ye, L. N. Pfeiffer and K. W. West, Physica E \textbf{22}, 104 (2004).
     \bibitem{hanfqhe}  Han Zhu, Yong P. Chen, P. Jiang, L. W. Engel, D. C. Tsui, L. N. Pfeiffer, and K. W. West
 Phys. Rev. Lett. {\bf 105}, 126803 (2010).


  \bibitem{bubres} R. M. Lewis,  P. D.  Ye, L. W. Engel, D. C. Tsui, L. Pfeiffer, and K. W. West,
   Phys. Rev. Lett. {\bf 89}, 136804   (2002).









\bibitem{flr}Hidetoshi Fukuyama and Patrick A. Lee, 
Phys. Rev. B {\bf 18}, 6245  (1978).
  \bibitem{chitra} R. Chitra, T. Giamarchi, and P. Le Doussal, 
Phys. Rev. Lett. 80, 3827 (1998); R. Chitra, T. Giamarchi, and P. Le Doussal,  Phys. Rev. B {\bf 65}, 035312 (2001).
 \bibitem{fertig} H. A. Fertig, Phys. Rev. B {\bf 59}, 2120 (1999).
\bibitem{foglerhuse} M. M. Fogler, and D. A. Huse, Phys. Rev. B  {\bf 62}, 7553 (2000).


  \bibitem{wanlireentrant} Wanli Li,  D. R. Luhman, D. C. Tsui, L. N. Pfeiffer, and K. W. West,
Phys. Rev. Lett. {\bf 105}, 076803 (2010).
 \bibitem{wanliscatt}Wanli Li, G. A. Cs\'{a}thy, D. C. Tsui, L. N. Pfeiffer, and K. W. West, Appl. Phys. Lett. {\bf 83}, 2832 (2003).
\bibitem{wanlisclg}
Wanli Li, G. A. Cs\'{a}thy, D. C. Tsui, L. N. Pfeiffer, and K. W. West,
Phys. Rev. Lett. {\bf 94}, 206807 (2005).
\bibitem{wanlisclganderson}
Wanli Li, C. L. Vicente, J. S. Xia, W. Pan, D. C. Tsui, L. N. Pfeiffer, and K. W. West,
Phys. Rev. Lett. {\bf 102}, 216801 (2009).
\bibitem{wanlisclgcross}
Wanli Li, J. S. Xia, C. Vicente, N. S. Sullivan, W. Pan, D. C. Tsui, L. N. Pfeiffer, and K. W. West,
Phys. Rev. B {\bf 81}, 033305 (2010).

  \bibitem{byounglonu} B.-H. Moon, L. W. Engel, D. C. Tsui, L. N. Pfeiffer, and K. W. West,  Phys. Rev. B {\bf 89}, 075310 (2014).

  
  \bibitem{zhwimbal} Zhihai Wang, Yong P. Chen, Han Zhu, L. W. Engel, D. C. Tsui, E. Tutuc, and M. Shayegan,
Phys. Rev. B {\bf 85}, 195408 (2012)
  \bibitem{aandm} N. W.  Ashcroft and N. D. Mermin, {\em Solid State Physics }, (Saunders, Philadelphia, 1976).
     \bibitem{barret}S. E. Barrett, G. Dabbagh, L. N. Pfeiffer, K. W. West, and R. Tycko, Phys. Rev. Lett \textbf{74}, 5112 (1995).
  \bibitem{sondhi}  S. L. Sondhi, A. Karlhede, S. A. Kivelson, and E. H. Rezayi, Phys. Rev. B \textbf{47}, 16419 (1993).
  \bibitem{schmeller} A. Schmeller \textsl{et al.}, Phys. Rev. Lett. \textbf{75}, 4290 (1995).
  \bibitem{bayotmelinte} V. Bayot, E. Grivei, S. Melinte, M. B. Santos, and M. Shayegan, Phys. Rev. Lett. 76, 4584 (1996); 79, 1718 (1997); S. Melinte,E. Grivei, V. Bayot, and M. Shayegan, ibid. 82, 2764 (1999).
\bibitem{fertigsk} R. C\^ot\'e, A. H. MacDonald, Luis Brey, H. A. Fertig, S. M. Girvin, and H. T. C. Stoof, Phys. Rev. Lett. \textbf{78}, 4825 (1997).
 
\bibitem{pricezhulouie}R. Price, Xuejun Zhu, P. M. Platzman, and Steven G. Louie,
Phys. Rev. B {\bf 48}, 11473 (1993).
 
  \bibitem{jainbook}  
J.~K. Jain,   \emph{Composite Fermions}.   Cambridge University Press, Cambridge, 2007.
 
  \bibitem{chitragiamarchiB0}  R. Chitra  and T. Giamarchi  Eur. Phys. J. B 44, 455Ð467 (2005).
 
  \bibitem{yac} S. Ilani,  J. Martin,  E. Teitelbaum,  J. H. Smet,  and D. Mahalu, Nature {\bf 427}, 328 (2004).
  \bibitem{jameikivspiv} 
Reza Jamei, Steven Kivelson, and Boris Spivak,
Phys. Rev. Lett. {\bf 94}, 056805 (2005). 
  \bibitem{kivspiv}  Boris Spivak and Steven A. Kivelson
Phys. Rev. B {\bf 70}, 155114 (2004).

\end{thebibliography}
\end{document}